\documentstyle[prd,aps,multicol,psfig]{revtex}

\draft

\renewcommand{\narrowtext}{\begin{multicols}{2} \global\columnwidth20.5pc}
\def\to{\rightarrow}

\def\AJ{{Astrophys.\ J.\ }}

\def\ASJ{{Astron.\ J.\ }}

\def\FP{{Fortsch.\ Phys.\ }}
\def\GRG{{Gen.\ Rel.\ Grav.\ }}

\def\JHEP{{JHEP} }

\def\NAT{{Nature (London)} }

\def\NC{{Nuovo Cim.\ }}
\def\NP{{Nucl.\ Phys.\ }}
\def\PL{{Phys.\ Lett.\ }}
\def\PR{{Phys.\ Rev.\ }}
\def\PRD{{Phys.\ Rev.\ D} }
\def\PRL{{Phys.\ Rev.\ Lett.\ }}

\def\RMP{{Rev.\ Mod.\ Phys.\ }}

\def\al{\alpha}

\def\de{\delta}

\def\ve{\varepsilon}

\def\th{\theta}

\def\ka{\kappa}

\def\rh{\rho}

\def\si{\sigma}

\def\ta{\tau}

\def\De{\Delta}
\def\Th{\Theta}
\def\La{\Lambda}

\def\Om{\Omega}

\def\cl{{\cal L}}

\def\fr#1#2{{{#1} \over {#2}}}
\def\half{{\textstyle{1\over 2}}}

\def\frac#1#2{{\textstyle{{#1}\over {#2}}}}

\def\lsim{\mathrel{\rlap{\lower4pt\hbox{\hskip1pt$\sim$}}
    \raise1pt\hbox{$<$}}}
\def\gsim{\mathrel{\rlap{\lower4pt\hbox{\hskip1pt$\sim$}}
    \raise1pt\hbox{$>$}}}
\def\sqr#1#2{{\vcenter{\vbox{\hrule height.#2pt
         \hbox{\vrule width.#2pt height#1pt \kern#1pt
         \vrule width.#2pt}
         \hrule height.#2pt}}}}

\def\prt{\partial}
\def\lrpartial{\raise 1pt\hbox{$\stackrel\leftrightarrow\partial$}}

\def\etal{{\it et al.}}

\newcommand{\beq}{\begin{equation}}
\newcommand{\eeq}{\end{equation}}
\newcommand{\bea}{\begin{eqnarray}}
\newcommand{\eea}{\end{eqnarray}}
\newcommand{\rf}[1]{(\ref{#1})}

\begin{document}

\title{Cosmological acceleration, varying couplings, and Lorentz breaking}

\author{Orfeu Bertolami,$^a$
Ralf Lehnert,$^b$ Robertus Potting,$^b$ and Andr\'e Ribeiro$^a$}

\address{$^a$Departamento de F\'{\i}sica, Instituto Superior T\'ecnico,
1049-001 Lisboa, Portugal}
\address{$^b$CENTRA, \'Area Departamental de F\'{\i}sica,
Universidade do Algarve,
8000-117 Faro, Portugal}

\maketitle

\begin{abstract}
Many candidate fundamental theories
contain scalar fields
that can acquire spacetime-varying expectation values
in a cosmological context.
Such scalars typically obey
Lorentz-violating effective dispersion relations.
We illustrate this fact
within a simple supergravity model
that also exhibits the observed late-time cosmological acceleration
and implies varying electromagnetic couplings.
\end{abstract}


\narrowtext

\section{Introduction}

An important question in present-day cosmology
concerns the expansion history of our universe.
Recent measurements
indicating a late-time period of accelerated expansion \cite{Perlmutter}
have been met with great interest.
Possible theoretical explanations for this observation
typically involve new fundamental physics
including a cosmological constant \cite{Bento1},
quintessence-type models \cite{early}
with one \cite{quint1} or two \cite{quint2} scalar fields,
k-essence \cite{kessence},
and exotic equations of state
like that of the generalized Chaplygin gas \cite{Kamenshchik}.

Other astrophysical observations
claim evidence for a time-dependent fine-structure parameter \cite{Webb}.
Early speculations in the subject of varying couplings
date back to Dirac's large-number hypothesis \cite{Dirac}.
Subsequent theoretical investigations have shown
that a spacetime dependence of the fine-structure parameter
arises naturally in candidate fundamental theories
and is often accompanied
by variations of other gauge or Yukawa couplings \cite{theo}.
In light of these facts,
a confirmation of the experimental observations
and the search for realistic models
that permit other particle-physics and cosmological predictions
have assumed particular urgency \cite{revs}.
In this context,
studies along the lines
of the Bekenstein model \cite{Bekenstein} and its generalizations
have suggested
dark matter and a cosmological constant \cite{Olive},
an ultra-light scalar field \cite{Gardner},
and quintessence \cite{Anchordoqui}
as driving entities
for a varying fine-structure parameter.

The presence of varying couplings
implies a breaking of invariance
under temporal or spatial translations.
This can be seen
as a special case of the violation
of spacetime symmetries,
which also include Lorentz and CPT invariance.
In fact,
time-dependent couplings
typically affect
these additional symmtries as well \cite{klp03}.
Note also
that Lorentz and CPT breakdown
has been suggested
in a variety of approaches to fundamental physics.
We mention
string theory \cite{kps},
spacetime foam \cite{ell98,suv},
nontrivial spacetime topology \cite{klink},
loop quantum gravity \cite{amu},
and noncommutative geometry \cite{chklo}.
Lorentz and CPT violation
provides therefore
another independent signature for an underlying theory.

Scalar fields are a common feature
in all of the above contexts.
This shows
that scalars play a key role
in the search for fundamental physics.
It becomes therefore interesting
to investigate cosmological expansion,
varying couplings, and Lorentz violation
within a single candidate fundamental model
containing scalars.
In the present work,
we shall consider
an $N=4$ supergravity model in four spacetime dimension.
The model contains two scalar fields,
an axion and a dilaton,
with a potential that we model with mass-type terms.
This framework,
although not fully realistic in its details,
incorporates many features
expected to be present in an encompassing theory:
the pure four-dimensional $N=4$ supergravity
is a limit of the $N=1$ supergravity in 11 dimensions,
which is contained in M theory.

The paper is organized as follows.
Section \ref{sec2}
describes the basics of our model.
In Sec.\ \ref{sec3},
we demonstrate how,
for a suitable range of parameters,
a late-time period of accelerated cosmological expansion
can arise in such a supergravity model.
The associated variation of the fine-structure parameter
and the electromagnetic $\th$ angle
is investigated in Sec.\ \ref{sec4}.
Section \ref{sec5}
discusses the violation of Lorentz and CPT symmetry
in the presence of spacetime-dependent scalars.
A brief summary is contained in Sec.\ \ref{sec6}.

\section{Basics}
\label{sec2}

The starting point for our investigation
is an isotropic homogeneous
flat $(k=0)$ Friedmann-Robertson-Walker universe
with the usual line element
\beq
ds^2 = dt^2 - a^2(t)\; (dx^2 + dy^2 + dz^2)\; ,
\label{frw}
\eeq
where $a(t)$ denotes the scale factor
and $t$ the comoving time.
The assumption of flatness
simplifies the analysis
and seems justified in light of recent measurements \cite{flat}.

In the present context,
we are interested in physical effects
past the formation of Hydrogen.
We therefore neglect the energy-momentum contribution
of radiation to the source of the Einstein equations.
Galaxies and other matter are modeled in a standard way
by the energy-momentum tensor of dust:
\beq
T_{\mu\nu} = \rh u_\mu u_\nu \; .
\label{dust}
\eeq
Here,
$\rh$ is the energy density of the matter
and $u^\mu$ is a unit timelike vector
orthogonal to the spatial hypersurfaces,
as usual.

In addition,
our model contains two real scalar fields,
an axion $A$ and a dilaton $B$,
coupled to the electromagnetic field $F_{\mu\nu}$.
The fields are described
by the Lagrangian density $\cl_{\rm sg}$ obeying
\bea
\ka \cl_{\rm sg}
&=&
-\frac 1 2 \sqrt{g} R
+\sqrt{g} ({\prt_\mu A\prt^\mu A + \prt_\mu B\prt^\mu B})/{4B^2}
\nonumber\\
&&
\qquad\!
-\frac 1 4 \ka \sqrt{g} M F_{\mu\nu} F^{\mu\nu}
-\frac 1 4 \ka \sqrt{g} N F_{\mu\nu} \tilde{F}^{\mu\nu}\; ,
\label{lag2}
\eea
where
\bea
M &=& \fr
{B (A^2 + B^2 + 1)}
{(1+A^2 + B^2)^2 - 4 A^2}\; ,
\nonumber\\
N &=& \fr
{A (A^2 + B^2 - 1)}
{(1+A^2 + B^2)^2 - 4 A^2}\; .
\label{N}
\eea
Here, $\tilde{F}^{\mu\nu}=\ve^{\mu\nu\rh\si}F_{\rh\si}/2$
denotes the dual field-strength tensor
and $g=-\det (g_{\mu\nu})$.
For convenience,
we rescale $\rh\to\rh/\ka$
and $F^{\mu\nu}\to F^{\mu\nu}/\sqrt{\ka}$
in what follows.
Then,
the gravitational coupling
$\ka$
does not appear explicitly in the equations of motion.
Under the assumption
that the dust arises through fermions
uncoupled from the scalars $A$ and $B$
and with the identification of $F^{\mu\nu}$
with one of the graviphotons,
this model fits into the framework
of $N=4$ supergravity in four dimensions \cite{cj,klp03}.

The pure N=4 supergravity
is known to be incomplete.
For example,
its spectrum differs
from the observed particle species in nature,
and the matter couplings of $F^{\mu\nu}$ are nonminimal.
This latter issue can be avoided
by gauging the internal SO(4) symmetry
leading to a potential for the scalars
that is unbounded from below \cite{dfr77}.
However,
besides additional fields,
any realistic situation
requires a stable vacuum.
We are therefore led to a phenomenological approach
and take the effective potential
to be bounded from below.
We further assume
that the scalars remain effectively uncoupled from the matter.
This may explicitly break the supersymmetry,
but does not represent a problem phenomenologically.

To lowest order,
we model the potentials for these scalars
by quadratic self-interactions
for the axion and dilaton
parametrized by the coefficients $m_A$ and $m_B$,
respectively.
Then,
our full Lagrangian density $\cl$ reads:
\beq
\cl=\cl_{\rm sg}-\half \sqrt{g} (m_A^2 A^2+ m_B^2 B^2)
+g_{\mu\nu}T^{\mu\nu}\; .
\label{lagr}
\eeq
Note that $m_A$ and $m_B$
cannot be identified directly with the masses of $A$ and $B$
because of the non-canonical kinetic terms.

The next step is
to determine the equations governing the evolution of the universe.
Our above assumptions of matter dominance, homogeneity, and isotropy imply
that we can neglect electromagnetic effects
for the large-scale dynamics of the universe.
We therefore set $F_{\mu\nu}=0$ for the moment.
Moreover,
the matter density $\rh$
and the scalars $A$ and $B$ are
(like the scale factor $a$)
functions of the comoving time $t$ only.
Then,
the respective $00$ and $jj$ components
of the Einstein equations
including the dust
are given by:
\bea
6\fr{\dot{a}^2}{a^2}&=&m_A^2 A^2+ m_B^2 B^2
+\fr{\dot{A}^2+\dot{B}^2}{2B^2}+2\rh\; ,\nonumber\\
4\fr{\ddot{a}}{a}+2\fr{\dot{a}^2}{a^2}&=&
m_A^2 A^2+ m_B^2 B^2
-\fr{\dot{A}^2+\dot{B}^2}{2B^2}\; ,\qquad
\label{eineq}
\eea
where the dot denotes a derivative
with respect to the comoving time.
The off-diagonal components are trivial.

The equations for the time evolution
of the axion and the dilaton
are obtained by variation of $\cl$
with respect to $A$ and $B$, as usual:
\bea
\fr d {dt}
\left(
\fr {a^3 \dot A}{B^2}\right)&=&
-2a^3m_A^2 A  \; ,\nonumber\\
\quad
\fr d {dt}
\left(
\fr {a^3 \dot B}{B^2} \right)&=&
-2a^3m_B^2 B - \fr {a^3 }{B^3} (\dot A^2 + \dot B^2)\; .
\label{abeq}
\eea

The above equations of motion for $A$ and $B$
imply that the energy contained in the scalars
is conserved separately:
\bea
0 &=& a^3D_{\mu}\Th^{\mu 0}\nonumber\\
&=& \fr d {dt}\left[\fr{a^3}{4B^2}(\dot{A}^2+\dot{B}^2)
+\fr{a^3}{2}(m_A^2 A^2+ m_B^2 B^2)\right]\nonumber\\
&&{}+\fr{3a^2\dot{a}}{2}
\left[\fr{1}{2B^2}(\dot{A}^2+\dot{B}^2)
-m_A^2 A^2-m_B^2 B^2
\right] ,
\label{encons}
\eea
where $\Th^{\mu\nu}$ denotes the symmetric energy-momentum tensor
associated with the scalars $A$ and $B$.
This essentially arises as a result of the fact
that our model fails to contain
interactions of the axion and the dilaton
with the fermion dust.
This is a phenomenologically reasonable assumption:
energy exchange between the scalars and the dust
can only be mediated by the electromagnetic field.
However,
the matter content of the universe
is electrically neutral and stable against photodecay
on macroscopic scales,
which excludes the required rates of photon exchange.

As an immediate consequence of the above result,
the energy-momentum tensor $T_{\mu\nu}$ of the dust
is also covariantly conserved.
This gives the equation
${d(\rh a^3)}/{dt} = 0$,
which can be integrated to yield
$\rh(t) = c_{\rm n}/a^3(t)$.
Here, the integration constant $c_{\rm n} = \rh_{\rm n} a_{\rm n}^3$
is determined by the matter density $\rh_{\rm n}=\rh(t_{\rm n})$
and scale size $a_{\rm n} = a(t_{\rm n})$ of the universe
at the present time $t_{\rm n}$.

Note that Eqs.\ \rf{eineq} and \rf{abeq}
are four equations
for the three unknown functions $a(t)$, $A(t)$, and $B(t)$.
To demonstrate the dependency among them,
one can proceed as follows.
Multiplication of the first Einstein equation in \rf{eineq} with $a^3$
and subsequent application of a time derivative
eliminates the term containing $\rh$ by energy conservation.
Moreover,
with the aid of the equations of motion for the scalars
in the form of Eq.\ \rf{encons},
one can rearrange the remaining expression
to obtain the second one of the  Einstein equations in \rf{eineq}.

\section{Accelerated cosmological expansion}
\label{sec3}

In this section,
we investigate the solutions
of the Eqs.\ \rf{eineq} and \rf{abeq},
which govern the dynamics of our model.
The two Einstein equations \rf{eineq} imply
\beq
6\fr{\ddot{a}}{a}=
m_A^2 A^2+m_B^2 B^2
-\fr{\dot{A}^2+\dot{B}^2}{B^2}-\rh\; .
\label{eineq2}
\eeq
Thus, in a physical model
at least one one of the parameters $m_A$ or $m_B$
must be nonzero
to yield an accelerated expansion rate $\ddot{a}(t)>0$,
as expected.
This is also consistent with a previous analysis
taking $m_A=m_B=0$:
in this case,
all equations of motion
can be integrated analytically
leading to the conventional decelerated expansion
$\sim t^{2/3}$ of a matter-dominated universe
in the present cosmological epoch \cite{klp03}.

In special situations
in which $m_A$, or $m_B$, or both are nonzero,
some analytical results can be obtained.
These results together with some general comments
on the spectrum of solutions
can be found in Appendix A.
However,
for the present purposes
numerical methods appear most promising for further progress.
A phenomenologically interesting input-parameter set
should minimize the deviation from experimental data.
As a result of the nonlinearity and the complexity
of the present system of differential equations,
it appears difficult to find a global minimum.
A numerical search
has given a variety of parameter sets
optimizing locally the departure from observations.
One such set
sufficient for our purposes
is the following:
\bea
m_A & = & 2.7688 \times 10^{-42}\, {\rm GeV}\; ,\nonumber\\
m_B & = & 3.9765 \times 10^{-41}\, {\rm GeV}\; ,\nonumber\\
c_{\rm n} & = & 2.2790 \times 10^{-84}\, {\rm GeV}^2\; ,\nonumber\\
a(t_{\rm n}) & = & 1 \; ,\nonumber\\
A(t_{\rm n}) & = & 1.0220426 \; ,\nonumber\\
\dot{A}(t_{\rm n}) & = & -8.06401\times 10^{-46}\, {\rm GeV}\; ,\nonumber\\
B(t_{\rm n}) & = & 0.016598 \; ,\nonumber\\
\dot{B}(t_{\rm n}) & = & -2.89477\times 10^{-45}\, {\rm GeV}\; .
\label{values}
\eea
The numerical analysis also showed
that there are two sets of solutions:
one in which the scale factor $a(t)$ increases with time
and another one in which the scale factor decreases.
In what follows,
we only consider the observationally relevant solution
with $\dot{a}(t)>0$.

Our next step
is to compare the solution determined by the parameters \rf{values}
with experimental data.
Measurements show
that the expansion of our universe
is consistent with a flat ``canonical'' model
containing nonrelativistic matter (dust)
and a cosmological constant $\La$.
The parameters for this canonical model
are inferred from observations
of high-redshift supernovae \cite{perl03}:
\bea
\Om_{\rm M}&=&0.30\pm 0.04\; ,\nonumber\\
\Om_\La&=&0.70\pm 0.04\; ,\nonumber\\
H_0&=&(70\pm 4)\:{\rm km\!\cdot\! s^{-1}\!\cdot\! Mpc^{-1}}\; .
\label{canvalues}
\eea
Here, $\Om_{\rm M}$ and $\Om_\La$ are the respective energy densities
associated with the matter and the $\La$ term
relative to the critical density $3{H_0}^2/\ka$.
The Hubble constant
is denoted by $H_0$,
as usual.
To make contact with observations,
we can therefore compare our solution
determined by the input values \rf{values}
with the above canonical model \rf{canvalues}.
To allow only small deviations of the supergravity cosmology
from the best-fit canonical model,
we have chosen the variations in the parameters \rf{canvalues}
to be somewhat smaller
than the experimental errors \cite{perl03}.
We also take the variations of the relative energy densities
$\Om_{\rm M}$ and $\Om_\La$
to be constrained by $\Om_{\rm M}+\Om_\La=1$,
so that the canonical model yields a flat ($k=0$) universe
for all parameters \rf{canvalues}.

The supergravity cosmology
leads to a value for the Hubble constant
of $H_0\simeq 69\, {\rm km\!\cdot\! s^{-1}\!\cdot\! Mpc^{-1}}$
consistent with observations.
The canonical model
yields $t_{\rm n}=1.35\times 10^{10}\, {\rm yr}$
for the present age of the universe.
Our supergravity cosmology
implies a similar age
of ${t_{\rm n}}=1.30\times 10^{10}\, {\rm yr}$.

\begin{figure}
\centerline{\psfig{figure=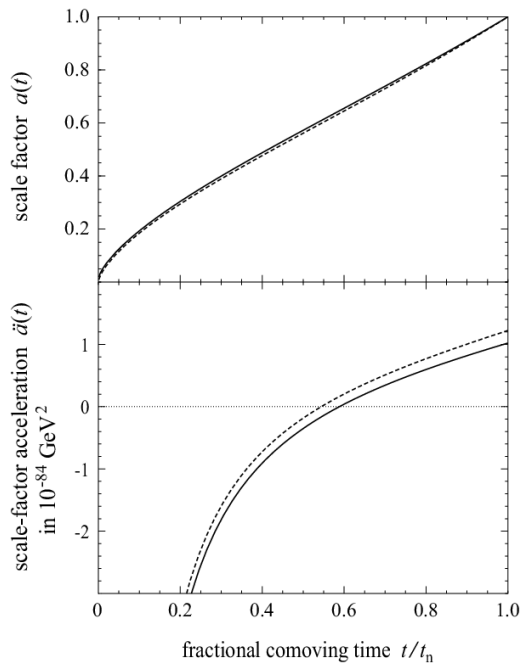,width=0.8\hsize}}
\smallskip
\caption{
Time evolution of the scale factor $a(t)$
and its second derivative $\ddot{a}(t)$.
The solid  and dashed lines
correspond to our supergravity universe
and the canonical model, respectively.
Note that for approximately the second half of its lifetime,
the expansion of the universe is speeding up
in both models.
}
\label{fig1}
\end{figure}

Other features of the cosmological expansion
are illustrated in Fig.\ \ref{fig1}.
The upper panel
shows the evolution of scale factor
from the Big Bang to the present.
The time dependence
of the scale-factor acceleration $\ddot{a}(t)$
is depicted in the lower panel.
The solid and the dashed lines refer
to our supergravity cosmology
and the canonical model, respectively.
The fractional comoving time
is determined by the corresponding
$t_{\rm n}$ values of each model
given in the previous paragraph.
Note the recent accelerated expansion of the universe
in both cosmologies
at comparable rates.

\begin{figure}
\centerline{\psfig{figure=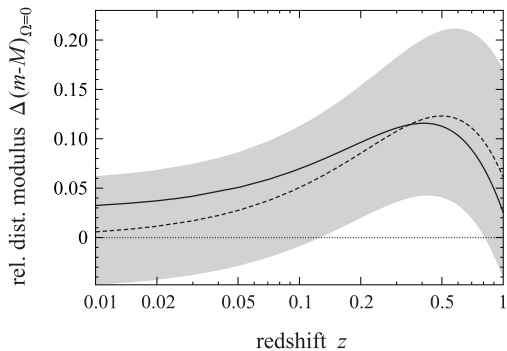,width=0.8\hsize}}
\smallskip
\caption{
Distance modulus
relative to an empty universe $\De(m-M)_{\Om=0}$
versus redshift $z$.
Our supergravity cosmology
is represented by the solid line
and the canonical model
by the dashed line.
The dotted line
corresponds to the empty universe.
The shaded region marks the
parameter range \rf{canvalues}.
}
\label{fig2}
\end{figure}

Although the differences in scale-factor evolution
bet{\-}ween our and the canonical model seem relatively minor
in Fig.\ \ref{fig1},
it is necessary to investigate
whether these deviations
are such
that consistency with observations
is maintained.
Experimentally,
the time dependence of the scale factor
is inferred from luminosity and redshift measurements
of distant cosmological objects.
Conventionally,
the luminosity observations are translated
into a luminosity distance $D_L$
that can be expressed
as a distance-modulus value $(m-M)$.
In Fig.\ \ref{fig2},
$\De(m-M)_{\Om=0}=(m-M)-(m-M)_{\Om=0}$
is plotted as a function of redshift $z$.
To emphasize the acceleration effects,
we have subtracted the distance modulus
$(m-M)_{\Om=0}$
of an empty $(k=-1)$ non-accelerated universe
with $a(t)=H_0t$ and
$H_0=70\:{\rm km\!\cdot\! s^{-1}\!\cdot\! Mpc^{-1}}$.
As before,
the solid line corresponds to our supergravity model
and the dashed line to the canonical model.
The dotted line represents the empty universe.
The shaded area corresponds to the parameter space \rf{canvalues}.
In the shown redshift region,
which corresponds to the current observational range of $z\lsim 1$,
both models explain the supernovae data.
At higher redshifts $z>1$,
the models could in principle be distinguished
by future experiments of this type.

\begin{figure}
\centerline{\psfig{figure=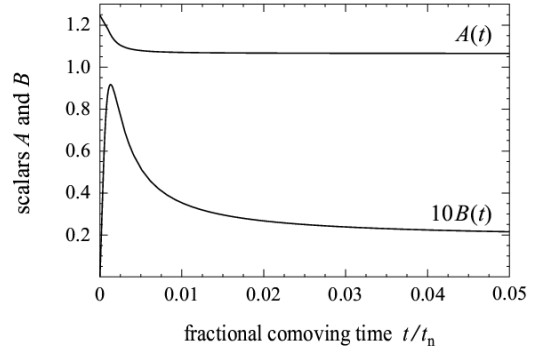,width=0.8\hsize}}
\smallskip
\caption{
Time evolution of the scalars $A(t)$ and $B(t)$
at early cosmological times.
In the recent past of our model universe,
which is not shown here,
$A(t)$ and $B(t)$ are essentially constant.
}
\label{fig3}
\end{figure}

Next,
we briefly discuss the time evolution
of the scalars $A(t)$ and $B(t)$.
The values \rf{values}
lead to the comoving-time dependence
of the axion and dilaton fields
depicted in Fig.\ \ref{fig3}.
For clarity,
the value of $B(t)$ has been multiplied
by a factor of ten.
The scalars $A(t)$ and $B(t)$ vary significantly only
in the early universe.
At later times,
they display a comparatively constant behavior,
which is phenomenologically necessary
because the axion and the dilaton
determine the variation of the fine-structure parameter,
which is tightly constrained experimentally.
A more detailed discussion of this topic
can be found in the next section.

\begin{figure}
\centerline{\psfig{figure=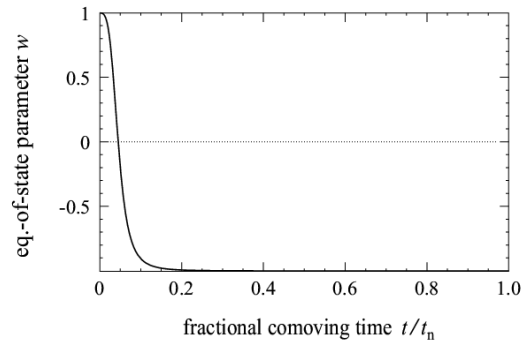,width=0.8\hsize}}
\smallskip
\caption{
Time evolution of the equation-of state parameter $w$.
At late times, $w\rightarrow -1$,
so that the scalars essentially
obey the cosmological-constant equation of state.
}
\label{fig4}
\end{figure}

We conclude this section
with a few comments
on two further quantities of interest.
One of them
is the equation-of-state parameter
$w\equiv p_{\rm s}/\rh_{\rm s}$,
where $p_{\rm s}$ denotes the pressure
of the scalars
and $\rh_{\rm s}$ their energy density.
Figure \ref{fig4} shows a plot of $w$
versus the fractional comoving time.
Note that $w$ is significantly different from $-1$
only at early times.
It follows
that in the recent past
the scalars affect the expansion of the universe
essentially in the same way as a cosmological constant.

\begin{figure}
\centerline{\psfig{figure=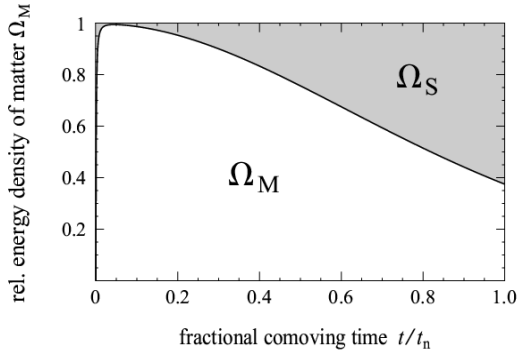,width=0.8\hsize}}
\smallskip
\caption{
Relative energy density of matter $\Om_{\rm M}$
versus fractional comoving time.
The shaded area shows  $\Om_{\rm S}$,
which corresponds
to the energy associated with the axion-dilaton background.
At late times,
$\Om_{\rm S}$ dominates,
which parallels the cosmological-constant situation.
}
\label{fig}
\end{figure}

The second quantity of interest
is the distribution of energy
between the nonrelativistic matter and the scalars.
In Fig.\ \ref{fig},
the relative energy density of the dust,
$\Om_{\rm M}$,
is plotted versus the fractional comoving time.
The relative energy density $\Om_{\rm S}$
associated with the scalars obeys $\Om_{\rm M}+\Om_{\rm S}=1$
in a flat universe,
so that the shaded region corresponds
to the time evolution of $\Om_{\rm S}$.
At late times,
$\Om_{\rm S}$ becomes the dominant energy component in our model.
This behavior,
which is characteristic for a conventional universe
with positive $\La$ energy,
arises
because the equation of state for the scalars
approaches that of a cosmological constant.
The supergravity cosmology
exhibits an energy distribution differing from the conventional case
only at early times:
$\Om_{\rm S}$ also dominates
close to the Big Bang.
However,
we remind the reader
that our simple model
neglects important effects
prior to the recombination time.

\section{Varying couplings}
\label{sec4}

In this section,
we consider excitations of $F_{\mu\nu}$
in the axion-dilaton cosmology
discussed above.
In most experimental investigations,
the spacetime regions involved
are small on a cosmological scale,
so that it is appropriate
to work in local inertial frames.
A few of the following results have been
derived previously for the case $m_A=m_B=0$ \cite{klp03}.
However,
we discuss them here anew,
both for completeness
and to emphasize qualitative differences
to the present case.

The conventional electrodynamics Lagrangian density in a local
inertial frame can be taken as
\beq
\cl_{\rm em} =
-\fr{1}{4 e^2} F_{\mu\nu}F^{\mu\nu}
- \fr{\th}{16\pi^2} F_{\mu\nu} \tilde{F}^{\mu\nu}\; .
\label{em}
\eeq
We remind the reader
that the $\th$-angle term
fails to contribute to the classical Maxwell equations
as long as $\th$ remains constant.
Comparison with our supergravity model yields
$e^2 \equiv 1/M$ and  $\th \equiv 4\pi^2 N$.
Note that $M$ and $N$
are functions of the scalars $A$ and $B$.
It follows
that the time dependence of $A$ and $B$
in the supergravity cosmology
induces a time variation of $e$ and $\th$.
Thus,
in the present model
the fine-structure parameter and the electromagnetic $\th$ angle
acquire related spacetime dependences
in a general coordinate system.

\begin{figure}
\centerline{\psfig{figure=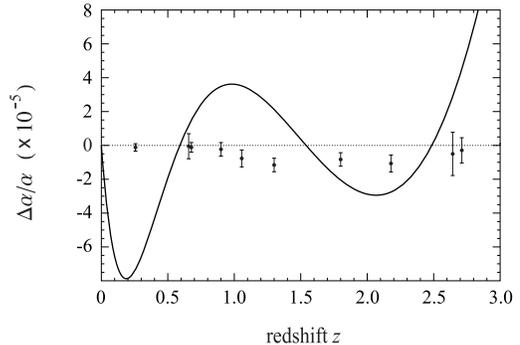,width=0.8\hsize}}
\smallskip
\caption{
Relative time variation of the electromagnetic coupling
versus fractional look-back time to the Big Bang.
The present model with the parameters \rf{values}
is represented by the solid line.
Also shown is the Webb dataset.
}
\label{fig5}
\end{figure}

In Fig.\ \ref{fig5},
the relative variation $(\al-\al_{\rm n})/\al_{\rm n}$
of the fine-structure parameter $\al=e^2/4\pi$
is plotted versus the fractional look-back time
$1-t/t_{\rm n}$ to the Big Bang.
Here, $\al_{\rm n}=1/137$ denotes the present value
of the electromagnetic coupling.
The solid line represents
our supergravity model with the input \rf{values}.
Also plotted is the recently reported Webb dataset \cite{Webb}
obtained from observations of high-redshift spectra.
Both the supergravity cosmology and the data
show a relative variation of $\al$ by parts in $10^{5}$.
Although our model fails to provide a good match to the data,
it does exhibit nonlinearities in $\al(z)$,
a necessary feature to fit all experimental constraints.

We mention that a generic choice of model parameters
typically yields a much larger variation of $\al$,
which can be understood as follows.
The supergravity cosmology
is governed by the coupled equations of motion
for the scale factor $a(t)$ and the scalars $A(t)$ and $B(t)$.
The observed value of the Hubble constant
provides the experimental constraint
$\dot{a}/a\sim  10^{-10}\;{\rm yr}^{-1}$.
One would therefore expect
that this is also the approximate variation
of the axion-dilaton background.
This value,
however,
is roughly six orders of magnitude larger
than the scale for $\dot{\al}/\al$
suggested by the Webb data \cite{Webb}.

In a general phenomenological model
involving some scalar field $\phi$,
the difference in the scales
for $\dot{a}/a$ and $\dot{\al}/\al$
is typically bridged as follows.
One normally expands $\al(\phi)=\al_{\rm n}(1+\xi \phi)+\ldots$,
which permits adjusting the parameter $\xi$
to obtain the desired order of magnitude for $\dot{\al}/\al$.
Note that this normally introduces an additional scale
into such a phenomenological model.
In the present supergravity framework,
however,
the dependence of $\al$
on the scalars $A$ and $B$ is fixed,
so that adjustable coefficients
in the electromagnetic coupling
are absent.
The above discussion suggests
that this is a critical issue
in models with parameter-free interactions
linking cosmological expansion and variation of couplings
via the same scalar field(s).

\section{Apparent Lorentz violation}
\label{sec5}

It is known
that spacetime-varying couplings
are typically associated
with Lorentz violation \cite{klp03}.
We summarize this result first
before further investigation.
In the present context,
Lorentz violation is exemplified
by the modified Maxwell equations
resulting from the effective Lagrangian density \rf{em}:
\beq
\fr{1}{e^2}\partial^{\mu}F_{\mu\nu}
-\fr{2}{e^3}(\partial^{\mu}e)F_{\mu\nu}
+\fr{1}{4\pi^2}(\partial^{\mu}\th)\tilde{F}_{\mu\nu}=0\; .
\label{Feom}
\eeq
For spacetime-independent $e$ and $\th$,
the usual electrodynamics equations are recovered.
In our axion-dilaton cosmology,
however,
the gradients of $e$ and $\th$
appearing Eq.\ \rf{Feom}
are nonzero,
approximately constant in local inertial coordinates,
and act as a nondynamical external background.
This vectorial background
determines a preferred direction
in the local inertial frame
violating particle Lorentz symmetry,
as defined in Refs.\ \cite{ck97,rl03}.

We remark
that general Lorentz and CPT violations of this type
are described by the Standard-Model Extension (SME)
\cite{ck97,ckp,kl01}.
This effective-field-theory framework
contains all coordinate-independent Lagrangian terms
formed by contracting Standard-Model operators
with Lorentz-violating parameters.
For example,
the term in Eq.\ \rf{Feom} involving $\th$
can be identified
with the $k_{AF}^{\mu}$ operator in the SME,
which breaks both Lorentz and CPT invariance.
We also mention
that the SME provides alternative explanations
for the baryon asymmetry in our universe \cite{baryo}
and for the observed neutrino oscillations \cite{neu}.

Particle Lorentz violation
through couplings varying on cosmological scales
is independent
of the chosen reference frame
and is not only a feature in electrodynamics.
Consider general equations of motions
in a given spacetime region.
If they contain the nonzero gradient of a coupling
in a particular set of local inertial coordinates,
the gradient is present in any coordinate system
associated with the spacetime region in question.
Although dynamical quantum scalar fields
are theoretically attractive
as driving entities for the variation of couplings,
the above argument for Lorentz violation
is also valid for classical scalar fields.
In fact,
the coupling need not be associated
with a dynamical field at all.

In what follows,
we shift our focus
from the electromagnetic sector
to the time-dependent scalars themselves.
In an effective vacuum
characterized by sizable cosmological solutions $A(t)$ and $B(t)$,
the propagation of the axion and the dilaton
is modified relative to the constant-background situation.
To investigate local effects of the background solutions
on the axion and the dilaton
at a certain spacetime point $x_0$
one can proceed in a local inertial frame.
Propagating disturbances in the axion-dilaton background
are described by small perturbations $\de A$ and $\de B$
away from the cosmological solution:
\bea
A(x)&\rightarrow& A(x)+\de A(x)\nonumber\\
B(x)&\rightarrow& B(x)+\de B(x)\; .
\label{pert}
\eea
The dynamics of the disturbances $\de A$ and $\de B$
is determined by the equations of motion \rf{abeq}.
In a general local inertial frame,
we obtain the following linearized equations:
\bea
0 &=& \Box \de A -2B^{\mu}\prt_{\mu}\de A
+2m_A^2B_{\rm b}^2\de A\nonumber\\
&&{}-2A^{\mu}\prt_{\mu}\de B
+(2A^{\mu}B_{\mu}+4m_A^2A_{\rm b}B_{\rm b})\de B\; ,\nonumber\\
0 &=& \Box \de B -2B^{\mu}\prt_{\mu}\de B
+2A^{\mu}\prt_{\mu}\de A\nonumber\\
&&{}+(6m_B^2B_{\rm b}^2-A^{\mu}A_{\mu}+B^{\mu}B_{\mu})\de B\; .
\label{ablineq}
\eea
Here,
the cosmological
axion-dilaton background
is described by $A_{\rm b}\equiv A(x)$,
$B_{\rm b}\equiv B(x)$,
$A^{\mu}\equiv B_{\rm b}^{-1}\prt^{\mu}A$,
and $B^{\mu}\equiv B_{\rm b}^{-1}\prt^{\mu}B$.
Note that $A_{\rm b}$ and $B_{\rm b}$
satisfy the equations of motion \rf{abeq}.
Locally,
the above definitions yield effectively constant quantities
because the axion-dilaton background
varies appreciably only on cosmological scales.
In what follows,
we can therefore take $A_{\rm b}$, $B_{\rm b}$, $A^{\mu}$, and $B^{\mu}$
as evaluated at $x=x_0$.

An ansatz with plane waves $\exp(-ip\cdot x)$
gives a set of algebraic momentum-space equations,
as ususal.
These two equations are governed
by the characteristic matrix $C$:
\beq
C=\left(\begin{array}{cc}
p^2-M_A^2-2iB\!\cdot\! p &
-2iA\!\cdot\! p
-2\tilde{M}^2\\
2iA\!\cdot\! p & p^2-M_B^2-2iB\!\cdot\! p
\end{array}\right)\; ,
\label{cmatrix}
\eeq
where $M_A^2\equiv 2m_A^2B_{\rm b}^2$,
$M_B^2\equiv 6m_B^2B_{\rm b}^2-A\!\cdot\! A+B\!\cdot\! B$,
and $\tilde{M}^2\equiv A\!\cdot\! B+2m_A^2A_{\rm b}B_{\rm b}$.
With the input values \rf{values},
$M_B^2$ and $\tilde{M}^2$ are positive
in the recent past of our model universe.
However,
other parameter choices
can lead to negative values for $M_B^2$ and $\tilde{M}^2$.
A detailed discussion of such a situation
would be interesting
but lies beyond the scope of the present work.

The plane-wave dispersion relation
describing the propagation of the various modes
is determined by $\det (C)=0$, as usual.
We obtain
\bea
0 & = & (p^2-M_A^2-2iB\!\cdot\! p)(p^2-M_B^2-2iB\!\cdot\! p)\nonumber\\
&&\qquad\qquad\qquad\qquad\qquad
{}+4(i\tilde{M}^2-A\!\cdot\! p)\;A\!\cdot\! p\;\; .
\label{disprel1}
\eea
The imaginary terms in this dispersion relation
are a direct consequence
of the non-Hermiticity of $C$.
They will lead to exponentially
increasing or decaying solutions.
This feature is characteristic
in cases with broken spacetime-translation symmetry
and the associated non-conservation of 4-momentum.

Another important feature of the dispersion relation \rf{disprel1}
are terms
in which the plane-wave 4-momentum $p^{\mu}$
is contracted with $A^{\mu}$ and $B^{\mu}$.
Since these 4-vectors
are taken as a constant nondynamical background
in the present context,
they select a direction in local inertial frames
violating particle Lorentz invariance.
As in the case of varying couplings,
this result is intuitively reasonable:
the effective vacuum containing the axion-dilaton background
acts as a spacetime-varying medium
in which the disturbances $\de A$ and $\de B$ propagate.
Moreover,
the spacetime-dependent cosmological solution for the scalars
breaks translation symmetry,
while translations and Lorentz transformations
are intertwined in the Poincar\'e group.
The above argument for Lorentz breaking
is not only confined
to the present supergravity cosmology.
It applies to any model
in which scalar fields acquire expectation values
varying on cosmological scales:
the propagation of localized disturbances
in the background expectation value
will typically violate Lorentz symmetry.
Note
that scalars with a ``rolling'' time dependence,
such as those in inflation, quintessence, and k-essence models,
play a key role in cosmology.

The tight experimental contraints
on the modification parameters in the SME \cite{cpt02} imply
that possible Lorentz and CPT violations must be minuscule
for known particles.
For instance,
the best bounds for radiation and matter
can be found in Refs.\ \cite{cfj,km} and \cite{bear},
respectively.
In the case of Lorentz-breaking dispersion relations,
kinematical threshold analyses are a useful tool \cite{rl03}.
Investigations of ultrahigh-energy cosmic rays,
for example,
can place Planck-scale bounds
on some SME coefficients for Lorentz violation \cite{bc00}.
The above experimental methods
are in principle also applicable in the present context.
However,
in the absence of direct observational evidence
for such fundamental scalars,
the present type of Lorentz violation
is phenomenologically less interesting
at the present time.
Moreover,
the discussion in the previous section implies
an expected size $\sim {\cal O} (H_0)$
for the background gradients,
which is below current experimental sensitivities.

Next,
we take advantage of coordinate Lorentz invariance
and continue our analysis
in a local comoving inertial frame.
This is also appropriate from an experimental viewpoint
because in realistic situations
laboratories move nonrelativistically
with respect to such frames.
Then, the scalars $A$ and $B$
depend only the time $t$ \cite{fn1},
so that $A_{\mu}=(\dot{A}, \vec{0})$ and
$B_{\mu}=(\dot{B}, \vec{0})$,
and the dispersion relation \rf{disprel1}
can be solved for $p^2=p^{\mu}p_{\mu}$:
\bea
p^2 & = & \half (M_A^2+M_B^2)+2i\dot{B}E\nonumber\\
&&\quad{}\pm\sqrt{\frac{1}{4}(M_A^2-M_B^2)^2
-4\dot{A}E(i\tilde{M}^2-\dot{A}E)} \; .
\label{disprel2}
\eea
Here,
we have denoted $p^0\equiv E$.

In principle,
the dispersion relation \rf{disprel2}
can be cast into the conventional form $E(\vec{p})$,
which involves in the present case
the general roots of a quartic equation.
However,
the expressions for the plane-wave energies
become more transparent
in certain limits.
For example,
we can consider a case
in which the non-standard dispersion-relation terms
are small, i.e.,
\beq
A_{\mu}A^{\mu},\:B_{\mu}B^{\mu},\:\tilde{M}^2\ll M_A^2,\: M_B^2\;.
\label{cond}
\eeq
Although the two scales are intertwined
in the equations of motion \rf{abeq},
this situation can in principle be realized.
Then,
in the nonrelativistic limit $|\vec{p}|\simeq 0$,
$E$ depends only on the above two scales,
and one expects $|E|\lsim {\cal O}(M_A,\: M_B,)$.
It follows that the dispersion relation \rf{disprel2}
reduces in leading order to
\beq
p^2 = \half (M_A^2+M_B^2)\pm \half (M_A^2-M_B^2)\; .
\label{nonrel}
\eeq
In such a situation,
$M_A$ and $M_B$ would therefore be
the effective axion and dilaton masses.
We remark
that the condition \rf{cond} is violated
in the model with the parameters \rf{values}.
This case is discussed below
when positivity and causality are investigated.

Another useful approximation
is the ultrarelativistic limit
\bea
E_+ & \simeq & \pm|\vec{p}|+A_0+iB_0\; ,\nonumber\\
E_- & \simeq & \pm|\vec{p}|-A_0+iB_0\; .
\label{ultrarel}
\eea
Here,
the subscripts $+$ and $-$
correspond to signs of the square root in Eq.\ \rf{disprel2}.
Asymptotically,
the real part of the energy variable grows linearly
with the 3-momentum,
as in the conventional case.
Note,
however,
the presence of imaginary terms.
Depending on the sign of $B_0$,
they lead to exponentially growing
or decaying solutions
implying non-conservation of 4-momentum.
This feature does not come as a surprise
because the cosmological axion-dilaton background
breaks spacetime-translation invariance.
One might argue
against exponentially growing solutions as being unphysical.
However,
it is important to observe
that ${\rm Im}(E)$ is determined
by the minuscule Lorentz-violating terms
and does not increase with the 3-momentum.
Moreover,
our solutions are valid only
within a localized spacetime region
with a characteristic size $\De t$
satisfying $\dot{B}\De t\ll 1$.
Since the local time $\ta$
obeys $\De \ta \lsim \De t$,
the change in amplitude
described by $\exp (\dot{B}\De\ta)\sim 1$,
and thus the resulting effect,
must be small.
One might also suspect
that the imaginary contributions to the energy
can lead to interpretational difficulty
when treated within quantum field theory (QFT).
For instance, the free QFT Hamiltonian
appears to be non-Hermitian
and Feynman boundary conditions for the Green function
may seemingly no longer be imposed.
However,
these apparent obstacles
can be avoided
if the time-dependent background
is treated perturbatively.
Then,
particle number,
for instance,
fails to be conserved,
which is consistent
with the above plane-wave solution of varying amplitude.

Lorentz-breaking dispersion relations
typically lead to positivity problems,
superluminal group velocities,
or both at high energies \cite{kl01,adam}.
In the present model,
the situation is further complicated
by the presence of imaginary terms,
so that we restrict ourselves to a few remarks.
As before,
we work in a local comoving inertial frame,
so that rotational symmetry is maintained.

\begin{figure}
\centerline{\psfig{figure=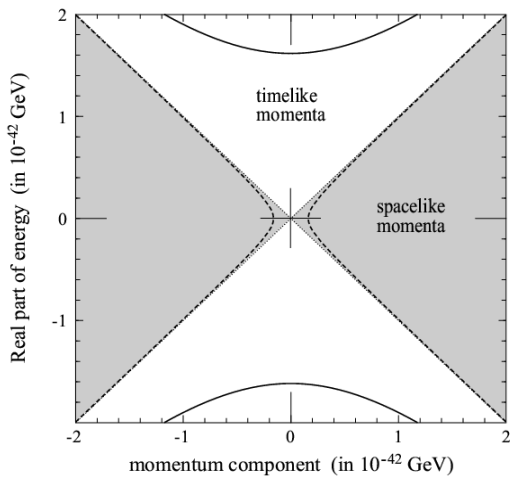,width=0.8\hsize}}
\smallskip
\caption{
Real part of the energy $E$
versus a momentum component
for the model parameters \rf{values}.
The dotted lines
correspond to the momentum-space lightcone.
The solid and dashed branches
are associated with the various roots
of the dispersion relation \rf{disprel1}.
Note the presence of spacelike momenta
and superluminal group velocities.
}
\label{figr}
\end{figure}

Figure \ref{figr} depicts the real part of the energy variable
as a function of a momentum component
in a model with parameters \rf{values}.
The dotted line represents the momentum-space lightcone.
As expected,
there are four branches,
each corresponding to one of the four roots
of the dispersion relation \rf{disprel1}.
Note that in the plotted momentum range
two of these branches,
shown as dashed lines,
lie in the shaded region of spacelike momenta.
The presence of spacelike 4-momenta
is normally associated with negative particle energies
in some inertial frames.
However,
instabilities are absent in the present model:
the conserved axion-dilaton energy
including the background remains always
real and positive definite,
as is evident from the energy-momentum tensor
appearing in Eq.\ \rf{eineq}.

\begin{figure}
\centerline{\psfig{figure=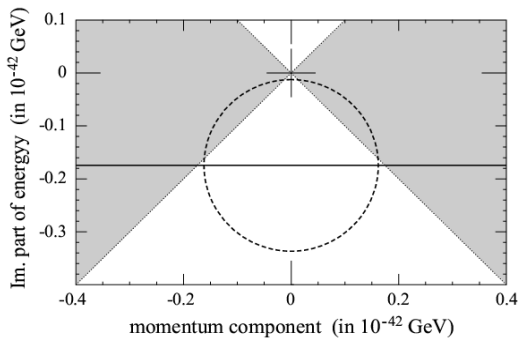,width=0.8\hsize}}
\smallskip
\caption{
Imaginary part of the energy $E$
versus a momentum component
for the model parameters \rf{values}.
The dotted lines
represent the momentum-space lightcone.
The solid and dashed lines
correspond to the respective solid and dashed branches
in Fig.\ \ref{figr}.
For large values of the momentum,
the solid and dashed lines lie on top of each other.
The region of spacelike momenta is shaded.
}
\label{figi}
\end{figure}

Asymptotically,
the magnitude of the group velocity
$\vec{v}_{\rm g}=\nabla_{\!\vec{p}}\:E(\vec{p})$
equals unity
because the conventional Lorentz-symmetric term
$(p^{\mu}p_{\mu})^2$
dominates the dispersion relation \rf{disprel1}
at high energies.
However,
inspection of the slope of a dashed branch in Fig.\ \ref{figr}
reveals the possibility that $|\vec{v}_{\rm g}|>c$
at low energies.
For a background that is fixed and fundamentally nondynamical,
such superluminal group velocities violate microcausality \cite{kl01}.
But for conventional dynamical backgrounds
leading to group velocities exceeding $c$,
causality is typically maintained.
Examples in ordinary electrodynamics
can be readily identified:
the occurrence of superluminal $\vec{v}_{\rm g}$
in amplifying media with inverted atomic populations
has been verified theoretically and experimentally \cite{chiao},
but causality violations are absent \cite{diener}.
The underlying Lorentz-covariant dynamics
of the our supergravity cosmology
leads therefore to the conjecture
that causality is maintained in the present context.
The apparent superluminal particle propagation
would then be an artifact of the approximations
employed in the derivation of the dispersion relation \rf{disprel1}.
A complete investigation of these issues
would be interesting
but lies well outside the scope of this work.

For completeness,
the imaginary part of the energy variable
as a function of momentum
is plotted in Fig.\ \ref{figi}.
The momentum-space lightcone is again represented
by the dotted line.
The shaded region is associated with spacelike 4-momenta.
The solid and dashed lines
correspond to the solid and dashed branches in Fig.\ \rf{figr}.
At large 3-momenta,
all four branches overlap
and ${\rm Im}(E)$ is bounded,
which is in agreement with Eq.\ \rf{ultrarel}.

\section{Summary}
\label{sec6}

This work has considered cosmologies
with scalar fields
that are motivated
in candidate fundamental theories.
In such a context,
the scalars typically acquire varying expectation values
that can be associated with an accelerated expansion of the universe,
varying couplings,
and Lorentz-violation.

More specifically,
we have investigated
an $N=4$ supergravity model in four dimensions.
In this framework,
standard plausible arguments
lead to a variation of the axion and the dilaton
on cosmological scales.
As a result,
the propagation of the axion and the dilaton
is governed by a Lorentz-violating effective dispersion relation.
We expect this feature to be generic
in models with scalars varying on cosmological scales.

The axion-dilaton background
also affects the expansion of the universe
and results in spacetime-dependent electromagnetic couplings.
Model parameters exist
that lead to a behavior of the scale factor
consistent
with the observed late-time cosmological expansion.
The variation of $\al$
implied by this parameter set
lies mostly outside experimental constraints.
However,
the time dependence of $\al$
is roughly in the order of magnitude
suggested by the Webb data
and displays desirable nonlinear features.

\section*{Acknowledgments}
This project was supported in part
by the Funda\c{c}\~ao para a Ci\^encia e a Tecnologia (Portugal)
under grants POCTI/1999/FIS/36285 and POCTI/FNU/49529/2002.
The work of R.L.\ and R.P.\
is partially funded
by the Centro Multidisciplinar de Astrof\'{\i}sica (CENTRA).
R.L.\ acknowledges the hospitality
extended to him by G.\ Soff
at the Institut f\"ur Theoretische Physik
of the Technische Universit\"at Dresden,
where part of this work was carried out.

\begin{appendix}

\section{Additional remarks regarding the spectrum of solutions}

We begin our considerations
with the observation
that for all nontrivial physical input values
the universe is always expanding
in our supergravity cosmology.
Suppose this were not the case.
Then, at some time $t=t_0$,
there must be a transition
from the presently observed cosmological expansion
to a subsequent epoch involving contraction.
This implies $\dot{a}(t_0)=0$ at the time $t_0$
by continuity of $\dot{a}(t)$.
Then, the first of the Einstein equations \rf{eineq}
yields at $t=t_0$
\beq
m_A^2 A^2+ m_B^2 B^2
+\fr{\dot{A}^2+\dot{B}^2}{2B^2}+2\rh=0\; ,
\label{indirect}
\eeq
which can only be satisfied in the trivial situation
of an empty universe with zero total energy density
or in unphysical cases
involving negative matter densities $\rh<0$.

We continue our discussion
by considering
special choices for the parameters $m_A$ and $m_B$.
The situation $m_A=0$ and $m_B=0$ has been discussed previously
in the literature \cite{klp03}.
Let us therefore focus first on the case
$m_A\neq0$ and $m_B=0$.
With this choice,
we find the following analytical solution:
\bea
a(t) & = & \sqrt[3\;]{3c_{\rm n}}\;t^{2/3}\; ,\nonumber\\
A(t) & = & \pm(m_A t)^{-1}\; ,\nonumber\\
B(t) & = & \pm(m_A t)^{-1}\; .
\label{sola}
\eea
Here,
the signs in the equations for $A(t)$ and $B(t)$
can be selected independently.
The absence of integration constants indicates
that more general solutions exist
(cf.\ model P2 below).
Note that the cosmological expansion implied by Eqs.\ \rf{sola}
is essentially that of a conventional matter-dominated universe,
which always decelerates.
The time dependence of the fine-structure parameter
is given by
\beq
\al(t)=\fr{1}{4\pi}\;\fr{4+m_A^4t^4}{2m_At+m_A^3t^3}\; .
\label{fsa}
\eeq
It follows that at late times,
$\al$ increases linearly with $t$.

Next,
we look at the case
$m_A=0$ and $m_B\neq0$.
One can verify that
\bea
a(t) & = & \sqrt[3\;]{\frac{6}{5}c_{\rm n}}\;t^{2/3}\; ,\nonumber\\
A(t) & = & A_0+A_3 t^{-3}\; ,\nonumber\\
B(t) & = & \pm\sqrt{2}(m_B t)^{-1}\; ,
\label{solb}
\eea
is an asymptotic $(t\rightarrow\infty)$ solution.
The integration constants $A_0$ and $A_3$
can be chosen freely.
This solution also describes
a non-accelerated cosmological expansion.
For $A_0\neq\pm1$,
the late-time behavior of $\al(t)$ is given by
\beq
\al(t)=\fr{1}{4\sqrt{2}\pi}\;\fr{(1-A_0^2)^2}{1+A_0^2}m_Bt\; .
\label{fsb1}
\eeq
In a situation with $A_0=\pm1$,
the fine-structure parameter decreases asymptotically
as follows:
\beq
\al(t)=(\sqrt{2}\pi m_Bt)^{-1}\; .
\label{fsb2}
\eeq
We also mention
that with the choice $A_3=0$,
Eqs.\ \rf{solb} become an exact solution.
In this case,
the time-evolution of the electromagnetic coupling
reads
\beq
\al(t)=\fr{1}{4\pi}\;\fr{4+4(1+A_0^2)m_B^2 t^2+(1-A_0^2)^2m_B^4 t^4}
{2\sqrt{2}m_Bt+\sqrt{2}(1+A_0^2)m_B^3t^3}\; .
\label{fsb3}
\eeq

\begin{figure}
\centerline{\psfig{figure=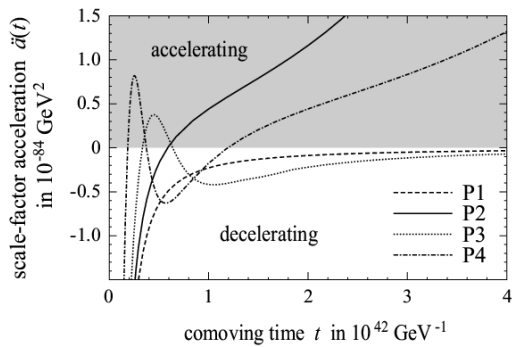,width=0.8\hsize}}
\smallskip
\caption{
Acceleration $\ddot{a}(t)$ of the scale factor
versus comoving time $t$
for the various input values given in Table I.
The shaded region corresponds to accelerated expansion.
The diversity in qualitative behavior is apparent.
}
\label{fig6}
\end{figure}

The general situation,
in which both $m_A$ and $m_B$ are nonzero,
seems to evade systematic analysis with analytical methods.
Numerical investigations
reveal the existence
of a broad spectrum of qualitatively different solutions.
Figure \ref{fig6}
conveys a flavor of the diversity in acceleration behavior
of our supergravity cosmology.
The second derivative of the scale factor,
$\ddot{a}(t)$,
is plotted versus the comoving time $t$
for the four parameter sets P1 to P4
given in Table I.
The shaded region marks accelerated cosmological expansion.
The input values P1
are associated with a cosmology of permanent deceleration,
qualitatively analogous
to a conventional matter-dominated universe
without cosmological constant.
The model parameters P2
lead to an initial deceleration period
followed by an asymptotic accelerated expansion.
This behavior agrees qualitatively
with that of the phenomenological model \rf{values}.
An overall decelerated expansion
with a transient period of acceleration
results from the input P3.
The parameter set P4
implies an initial deceleration,
followed by transient periods of acceleration and deceleration
before the expansion continues to accelerate.

It is worth emphasizing
that some of the solutions we have obtained
exhibit the interesting feature
of a transient accelerated expansion
(cf.\ model P3).
This is a possible fix for the recently discussed problem
concerning an eternally accelerated cosmological expansion.
Such universes would pose a challenge for string theory,
at least in its present formulation,
since string asymptotic states are inconsistent with spacetimes
that possess event horizons in the future \cite{Hellerman}.

\begin{center}
{\small TABLE I. Input-parameter sets P1, P2, P3, and P4.}
\vskip5pt
\begin{tabular}{lcccc}
\hline\hline
\multicolumn{1}{c}{\rule[-2mm]{0mm}{6mm}Parameter \rule{2mm}{0mm}}
& P1 & P2 & P3 & P4 \\ \hline
\rule{0mm}{5mm}$m_A$ in $10^{-42}\, {\rm GeV}$ \rule{2mm}{0mm}
& 0 & 1.5 & 0 & 1 \\
$m_B$ in $10^{-42}\, {\rm GeV}$ \rule{2mm}{0mm}
& 10 & 0 & 100 & 100 \\
$c_{\rm n}$ in $10^{-84}\, {\rm GeV}^2$ \rule{2mm}{0mm}
& 2 & 2 & 2 & 2 \\
$a(t_{\rm n})$ \rule{2mm}{0mm}
& 1 & 1 & 1 & 1 \\
$A(t_{\rm n})$ \rule{2mm}{0mm}
& 1.023 & 1.023 & 1.023 & 1.023 \\
$\dot{A}(t_{\rm n})$ in $10^{-47}\, {\rm GeV}$ \rule{2mm}{0mm}
& 47 & 47 & 47 & -100 \\
$B(t_{\rm n})$ \rule{2mm}{0mm}
& \rule{2mm}{0mm}0.022\rule{2mm}{0mm}
& \rule{2mm}{0mm}0.022\rule{2mm}{0mm}
& \rule{2mm}{0mm}0.022\rule{2mm}{0mm}
& \rule{2mm}{0mm}0.022 \\
$\dot{B}(t_{\rm n})$ in $10^{-45}\, {\rm GeV}$ \rule{2mm}{0mm}
& -25 & -25 & -25 & -60 \\
$t_{\rm n}$ in $10^{40}\, {\rm GeV}^{-1}$
\rule{2mm}{0mm}\rule[-2.5mm]{0mm}{1mm}& 56 & 51 & 54 & 51 \\ \hline\hline
\end{tabular}
\end{center}

\end{appendix}

\end{multicols}

\begin{references}
\bibitem{fn2}
There exist also parameter choices leading to a variation of $\al$
that approximately matches observations \cite{klp03,03rl1}. In
these cases, however, the resulting cosmological expansion is
phenomenologically uninteresting.
\bibitem{fn1}
In a local comoving inertial frame,
the time coordinate $\tilde{t}$
agrees to first order with comoving time $t$,
so that we can take $\tilde{t}\simeq t$.
\end{references}
\end{document}